\documentclass[a4paper,10pt]{article}
\usepackage{graphicx}
\usepackage{fancyheadings}
\begin{document}
\noindent
{\it 12th Hel.A.S Conference}\\
\noindent
{\it Thessaloniki, 28 June - 2 July, 2015}\\
\noindent
%
%
CONTRIBUTED LECTURE\\
\noindent
\underline{~~~~~~~~~~~~~~~~~~~~~~~~~~~~~~~~
~~~~~~~~}
\vskip 1cm
%
%
\begin{center}
{\Large\bf
Poetry in motion: Asteroseismology of $\delta$ Scuti Stars in Binaries using $Kepler$ Data
}
\vskip 0.5cm
%
%
{\it
A. Liakos$^1$ and P. Niarchos$^2$
}\\
%
%
$^1$National Observatory of Athens, Institute for Astronomy, Astrophysics, Space Applications and Remote Sensing, Penteli, Athens, Hellas\\
$^2$National \& Kapodistrian University of Athens, Department of Astrophysics, Astronomy and Mechanics, Zografos, Athens, Hellas
\end{center}
\vskip 0.7cm
%
%
\noindent
{\bf Abstract: }
The results of our six year systematic observational survey on candidate eclipsing binaries with a $\delta$~Sct component are briefly presented. A new catalogue for this kind of systems as well as the properties of their $\delta$~Sct members are also presented. The comparison between the components-pulsators and the single $\delta$~Sct stars shows that both the evolution and the pulsating properties differ significantly. Finally, we introduce the new era of studying stellar pulsations using high accuracy data from $Kepler$ mission and emphasizing the great opportunities that are now opened for a deep knowledge of the properties of stellar pulsations.


\section{Introduction}
Eclipsing binary systems (hereafter EBs) can be considered as extremely powerful tools for calculating the absolute parameters (i.e. masses, radii, temperatures, evolutionary status) of their members. Especially, when photometric analysis is combined with radial velocities curves analysis the results are become extremely accurate. Moreover, the $O-C$ analysis (i.e. Eclipse Timing Variation -- $ETV$) provides the means to study the EB's orbital period variations and connect them directly to various physical mechanisms (e.g. existence of tertiary component, mass transfer, mass loss etc).

Pulsating stars show light variations originated from intrinsic causes, such as pause of hydrostatic equilibrium. $\delta$ Scuti stars are pulsating stars that show both radial and non-radial pulsations. Pulsations are generated and preserved in the partial ionized zones inside the stellar interior and can be described quite well by the $\kappa$ and $\gamma$ mechanisms. Their mathematical description is based on the spherical harmonics.

Therefore, the study of pulsating stars in binary systems is quite useful for calculating the absolute parameters and the evolutionary status of pulsators. In addition, binarity effects (e.g. mass transfer, mass loss and tidal interactions) play an important role to the evolution and the oscillations of the pulsating stars-members of binaries, a fact which is a quite new subject of modern Astrophysics and is not well investigated so far. The first definition of this new subcategory of binaries was given by \cite{MK02,MK04} as $oEA$ (oscillating EA) stars. These systems are in semidetached configuration and they contain a (B)A-F spectral type, mass-accreting main-sequence pulsator. So far, many pulsating stars in binaries have been found \cite{ZH10}, but our study focuses on the cases of binaries with a pulsating star of $\delta$ Scuti type.

The history of these systems is relatively short. \cite{SO06a}, based on a sample of 20 systems, mentioned for the first time the possible existence of a connection between pulsation ($P_{\rm pul}$) and orbital ($P_{\rm orb}$) periods for EBs with a $\delta$~Sct member. \cite{SO06b} published two catalogues, one with confirmed systems of this type and a second with candidates. A new catalogue with 75 confirmed systems with a $\delta$~Sct member and new correlations between their fundamental stellar characteristics were later published by \cite{LI12}. \cite{ZH13} made the first theoretical attempt to correlate $P_{\rm pul}$-$P_{\rm orb}$. Finally, \cite{LI15} published a new catalogue containing 107 cases in total and noticed for the first time that there is a threshold of $P_{\rm orb}\sim13$~days below which $P_{\rm pul}$ and $P_{\rm orb}$ are strongly correlated, while above that these quantities seem to be independent.

In the present study we outline the results of our observational campaign, we present an updated catalogue and correlation for $P_{\rm pul}-P_{\rm orb}$, and finally we briefly present our method for finding new systems using $Kepler's$ mission databases.

\section{The observational campaign (2006-2012)}
We used the catalogue of \cite{SO06b} for selecting candidate systems including a $\delta$~Sct component. The campaign began in 2006 and lasted until 2012. Photometric observations were obtained using the telescopes (40~cm, 25~cm, 20~cm) of the Gerostathopoulion Observatory of the University of Athens and the 1.2~m telescope of the Kryonerion Astronomical Station of the National Observatory of Athens. Limited spectroscopic observations were made at Skinakas Observatory using the 1.3~m telescope. In total, we observed 108 candidate systems from which 13 were found to be new cases, while for other 8 we are not certain about their possible pulsational behaviour mostly due to instrumentation limits. Moreover, we observed systematically other 9 known oEA systems for better multiband light curve coverage in order to determine more accurately their pulsation properties. Summarizing the above, complete multi-colour light curves and systematic observations were made for 21 systems with a $\delta$~Sct component. The results of the individual systems as well as details for the methods of analyses we applied can be found in \cite{LI12,LI13,LI15}. Finally, the results are also available online\footnote{http://alexiosliakos.weebly.com/binaries-with-a-delta-sct-member.html}.

\section{Updated catalogue, statistical results and $P_{\rm pul}$-$P_{\rm orb}$ correlation}
\begin{figure}[h]
\centering
\begin{tabular}{cc}
\includegraphics[width=7.5cm]{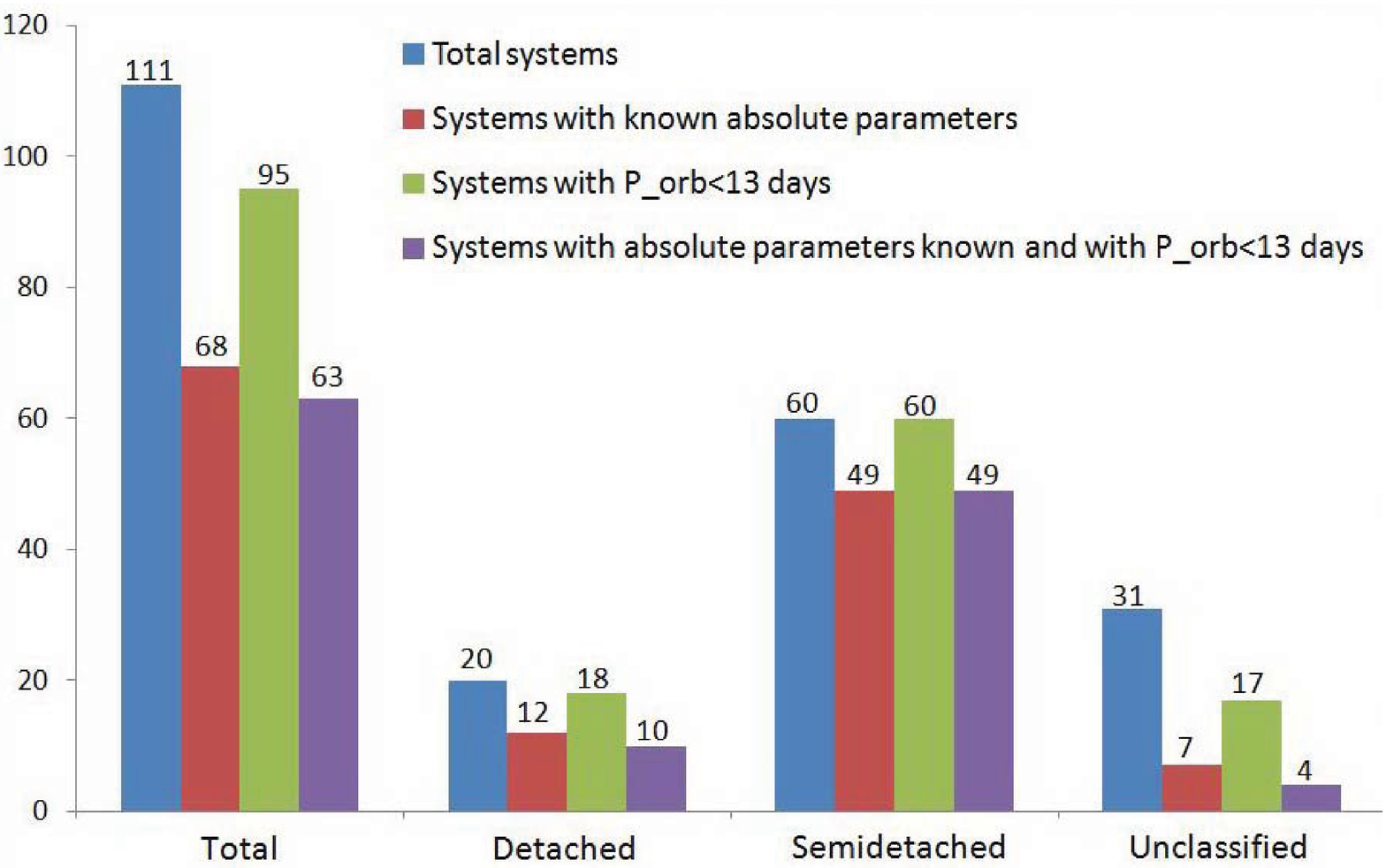}&\includegraphics[width=7.5cm]{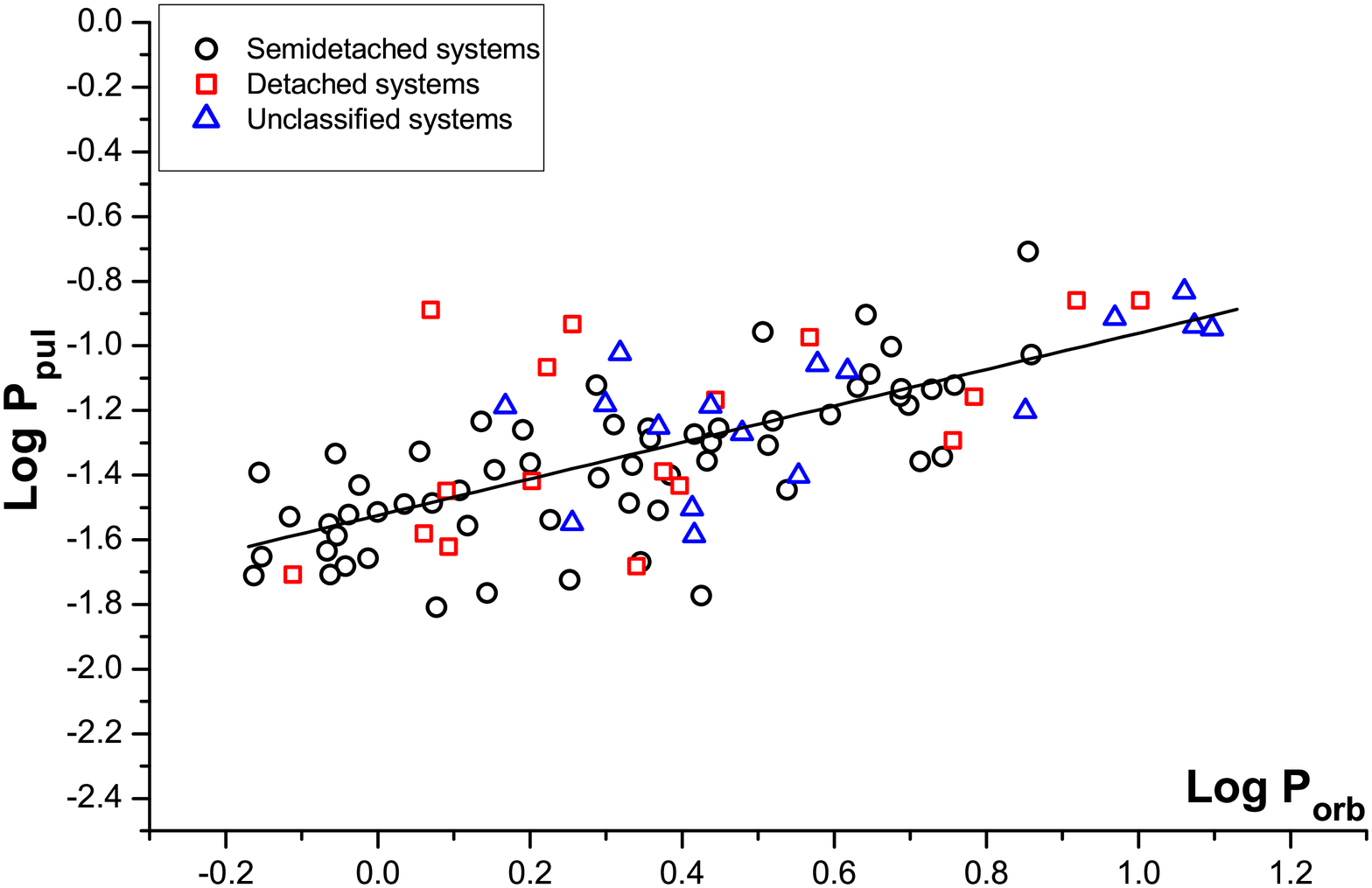}
\end{tabular}
\caption{Left: The statistics of the currently known binaries with a $\delta$~Sct component. Right: $P_{\rm pul}$-$P_{\rm orb}$ correlation for systems with $P_{\rm orb}<13$~days.}
\label{Fig1}
\end{figure}

In fig.~\ref{Fig1}-left the statistics for binaries with a $\delta$~Sct component is presented, while the linear fit on the $P_{\rm pul}$-$P_{\rm orb}$ data points for systems with $P_{\rm orb}<13$~days is given in fig.~\ref{Fig1}-right. The new catalogue of all the currently known systems (111 in total) is presented in Table \ref{Tab1} and includes the name and the orbital period ($P_{\rm orb}$) of the binary and the dominant pulsation frequency ($f_{\rm pul}$) of the $\delta$~Sct star. The catalogue is available online\footnote{http://alexiosliakos.weebly.com/catalogue.html} and is being updated systematically. The online version of the catalogue contains also the absolute parameters (i.e. mass, radius, effective temperature) of the $\delta$~Sct star, the geometrical type, the mass ratio and all the relevant references for each system.

The new correlation between orbital and pulsation periods, based on the sample of 95 systems with $P_{\rm orb}<13$~days, is the following:
\begin{equation}\label{}
\log P_{\rm puls}=0.56 (6) \log P_{\rm orb}-1.52(3)
\end{equation}

\begin{table}[h]
\centering
\caption{The updated catalogue of binaries with a $\delta$ Sct member}
\label{Tab1}
\scalebox{0.86}{
\begin{tabular}{lcc lcc lcc}
\hline
Name        &   $P_{\rm orb}$   & $f_{\rm pul}$&Name        &   $P_{\rm orb}$   & $f_{\rm pul}$&Name        &   $P_{\rm orb}$   & $f_{\rm pul}$\\
            &   [days]      &       [c/d]&            &   [days]      &       [c/d]&            &   [days]      &       [c/d]\\
\hline
\multicolumn{9}{c}{Systems with $P_{\rm orb}<$13 days}\\
\hline
 Aqr CZ	      &	0.86275	&	35.508	&	Eri AS	&	2.66410	&	59.172	&	Lep RR	&	0.91543	&	33.280	\\
Aqr DY	     &	2.15970	&	23.370	&	Eri TZ	&	2.60610	&	18.718	&	Lyn CL	&	1.58606	&	23.051	\\
Aql QY	     &	7.22954	&	10.656	&	Gru RS	&	11.5	&	6.803	&	Lyn CQ	&	12.50736	&	8.868	\\
Aql V0729	&	1.28191	&	28.034	&	GSC 3889-0202	&	2.71066	&	22.676	&	Mic VY	&	4.43637	&	12.234	\\
Aql V1464	&	0.69777	&	24.621	&	GSC 4293-0432	&	4.38440	&	8.000	&	Oph V0577	&	6.07910	&	14.388	\\
Aur KW (14)	&	3.78900	&	11.429	&	GSC 4588-0883	&	3.25855	&	20.284	&	Oph V2365 	&	4.86560	&	14.286	\\
Aur V0551	&	1.17320	&	7.727	&	HD 061199	&	3.57436	&	25.257	&	Ori FL	&	1.55098	&	18.178	\\
Boo EW	&	0.90630	&	48.008	&	HD 062571	&	3.20865	&	9.051	&	Ori FR	&	0.88316	&	38.600	\\
Boo YY	&	3.93307	&	16.318	&	HD 099612	&	2.77876	&	14.714	&	Ori V0392	&	0.65928	&	40.578	\\
Cam Y	&	3.30570	&	17.065	&	HD 172189	&	5.70165	&	19.608	&	Ori V1004	&	2.74050	&	15.365	\\
Cap TY	&	1.42346	&	24.222	&	HD 207651	&	1.47080	&	15.434	&	Pav MX	&	5.73084	&	13.227	\\
Cas AB	&	1.36690	&	17.153	&	HD 220687	&	1.59425	&	26.169	&	Peg BG	&	1.95267	&	25.544	\\
Cas IV	&	0.99852	&	32.692	&	Her BO	&	4.27281	&	13.430	&	Peg GX	&	2.34100	&	17.857	\\
Cas RZ	&	1.19530	&	64.197	&	Her CT	&	1.78640	&	52.937	&	Per AB	&	7.16030	&	5.107	\\
Cas V0389	&	2.49477	&	27.100	&	Her EF	&	4.72920	&	10.070	&	Per IU	&	0.85700	&	43.131	\\
Cep XX	&	2.33732	&	32.258	&	Her LT	&	1.08404	&	30.800	&	Pup HM	&	2.58972	&	31.900	\\
Cet WY	&	1.93969	&	13.211	&	Her TU	&	2.26690	&	17.986	&	Pyx XX	&	1.15000	&	38.110	\\
Cha RS	&	1.66987	&	11.628	&	Her V0644	&	11.85859	&	8.688	&	Ser AO	&	0.87930	&	21.505	\\
CMa R	&	1.13590	&	21.231	&	Her V0994	&	2.08309	&	10.563	&	Sge UZ	&	2.21574	&	46.652	\\
$^{a}$105906206	&	3.69457	&	9.417	&	HIP 7666	&	2.37232	&	24.450	&	Tau AC	&	2.04340	&	17.535	\\
Cyg UW	&	3.45078	&	27.841	&	Hor TT	&	2.60820	&	38.700	&	Tel IZ	&	4.88022	&	13.558	\\
Cyg V0346	&	2.74330	&	19.920	&	Hya AI	&	8.28970	&	7.246	&	Tri X	&	0.97151	&	45.455	\\
Cyg V0469	&	1.31250	&	35.971	&	Hya RX	&	2.28170	&	19.380	&	Tuc $\theta$	&	7.10360	&	15.946	\\
Del BW	&	2.42319	&	25.100	&	KIC 03858884	&	10.04860	&	7.231	&	UMa IO	&	5.52017	&	22.015	\\
Dra HL	&	0.94428	&	26.914	&	KIC 04544587	&	2.18909	&	48.022	&	UMa VV	&	0.68738	&	51.299	\\
Dra HN	&	1.80075	&	8.558	&	KIC 06629588	&	2.26447	&	13.396	&	UNSW-V-500	&	5.35048	&	13.624	\\
Dra HZ	&	0.77294	&	51.068	&	KIC 06669809	&	0.73374	&	32.564	&	$^{b}$0975-17281677	&	3.01550	&	18.702	\\
Dra OO 	&	1.23837	&	41.867	&	KIC 10661783	&	1.23136	&	28.135	&	$^{b}$1200-03937339	&	1.17962	&	30.668	\\
Dra SX	&	5.16957	&	22.742	&	Lac AU	&	1.39259	&	58.217	&	Vel AW	&	1.99245	&	15.200	\\
Dra TW	&	2.80690	&	17.986	&	Leo DG	&	4.14675	&	11.994	&	Vel BF	&	0.70400	&	44.940	\\
Dra TZ	&	0.86603	&	50.993	&	Leo WY	&	4.98578	&	15.267	&	Vul 18	&	9.31	&	8.230	\\
Dra WX	&	1.80186	&	35.468	&	Leo Y	&	1.68610	&	34.484	&		&		&		\\
\hline
\multicolumn{9}{c}{Systems with $P_{\rm orb}>$13 days}\\
\hline
Cas $\beta$	&	27	&	9.911	&	Hya KZ	&	9782	&	16.807	&	Tau $\varrho$ 	&	460.7	&	14.925	\\
CVn 4	&	124.44	&	8.595	&	KIC 11754974	&	343	&	16.342	&	Tau $\theta^{2}$ 	&	140.72816	&	13.228	\\
Del $\delta$	&	40.58	&	6.378	&	Lyn SZ	&	1181.1	&	8.299	&	Tau V0777 	&	5200	&	5.486	\\
Dra GK	&	16.96	&	8.790	&	Ori EY	&	16.78781	&	9.709	&	Vir FM	&	38.324	&	13.908	\\
HD 050870	&		&	17.162	&	Ori FO	&	18.80062	&	34.247	&		&		&		\\
HD 051844	&	33.49830	&	12.213	&	Peg IK	&	21.72400	&	22.727	&		&		&		\\
\hline
\multicolumn{9}{l}{$^{a}$CoRoT ID, $^{b}$USNO A2.0 ID }\\
\end{tabular}}
\end{table}

\section{Kepler's era}
The $Kepler$ mission has contributed a lot to asteroseismology and has been proved as a very powerful tool for obtaining long time-series of data. The main advantages of $Kepler's$ data for asteroseismic studies are: a) their high photometric accuracy (order of $10^{-4}$~mag), b) their high time resolution (short cadence data have $\sim1$~min resolution), and c) the continuous recording (no time gaps), which is very critical for eliminating alias effects during the frequency search. In addition, there is an excellent open database\footnote{http://keplerebs.villanova.edu/} for eclipsing binaries, created by \cite{PR11}, which can be easily used for searching, finding and downloading instantly data for further analysis.

\begin{figure}[h]
\centering
\begin{tabular}{cc}
\includegraphics[width=7.5cm]{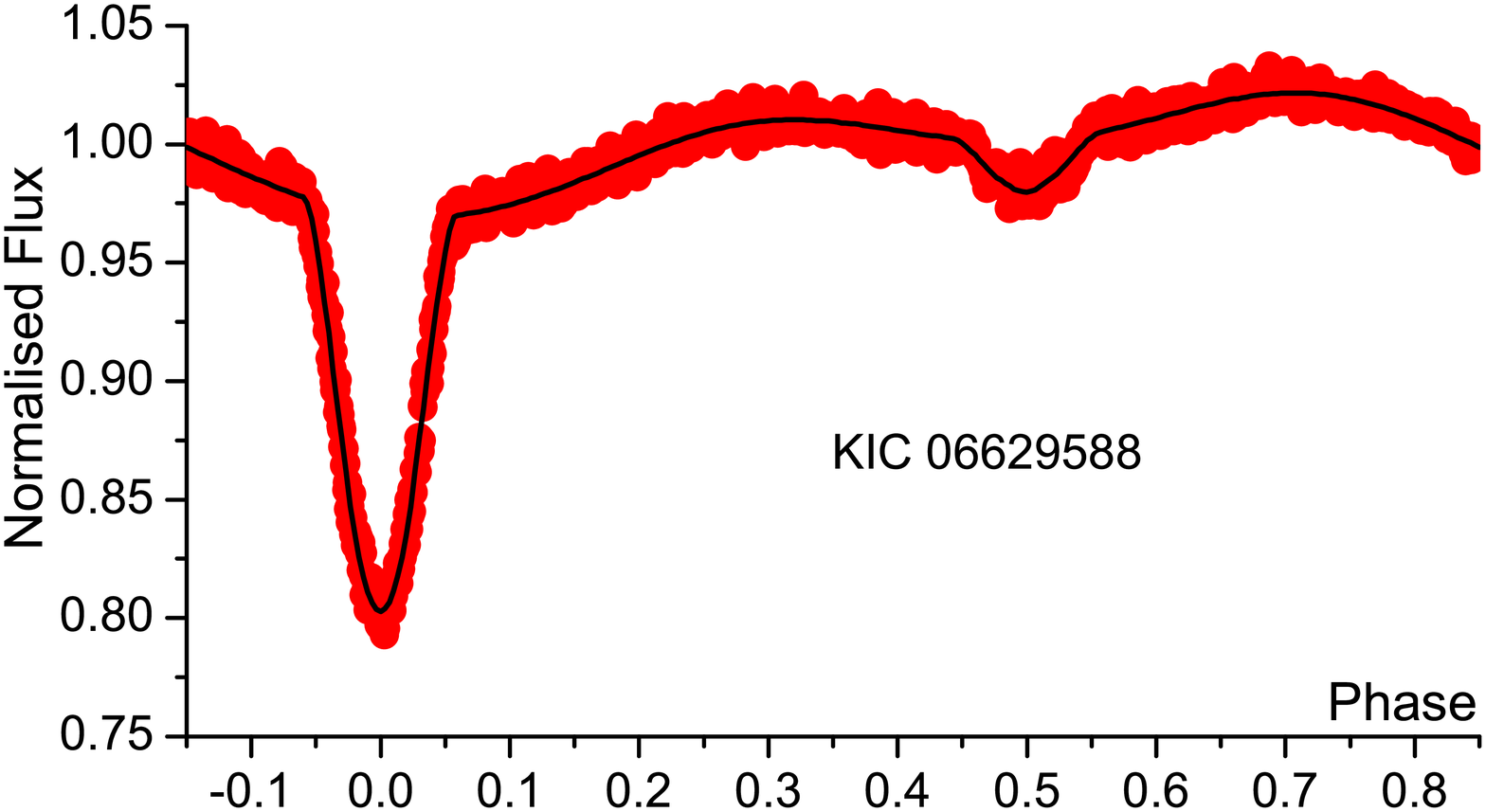}&\includegraphics[width=7.5cm]{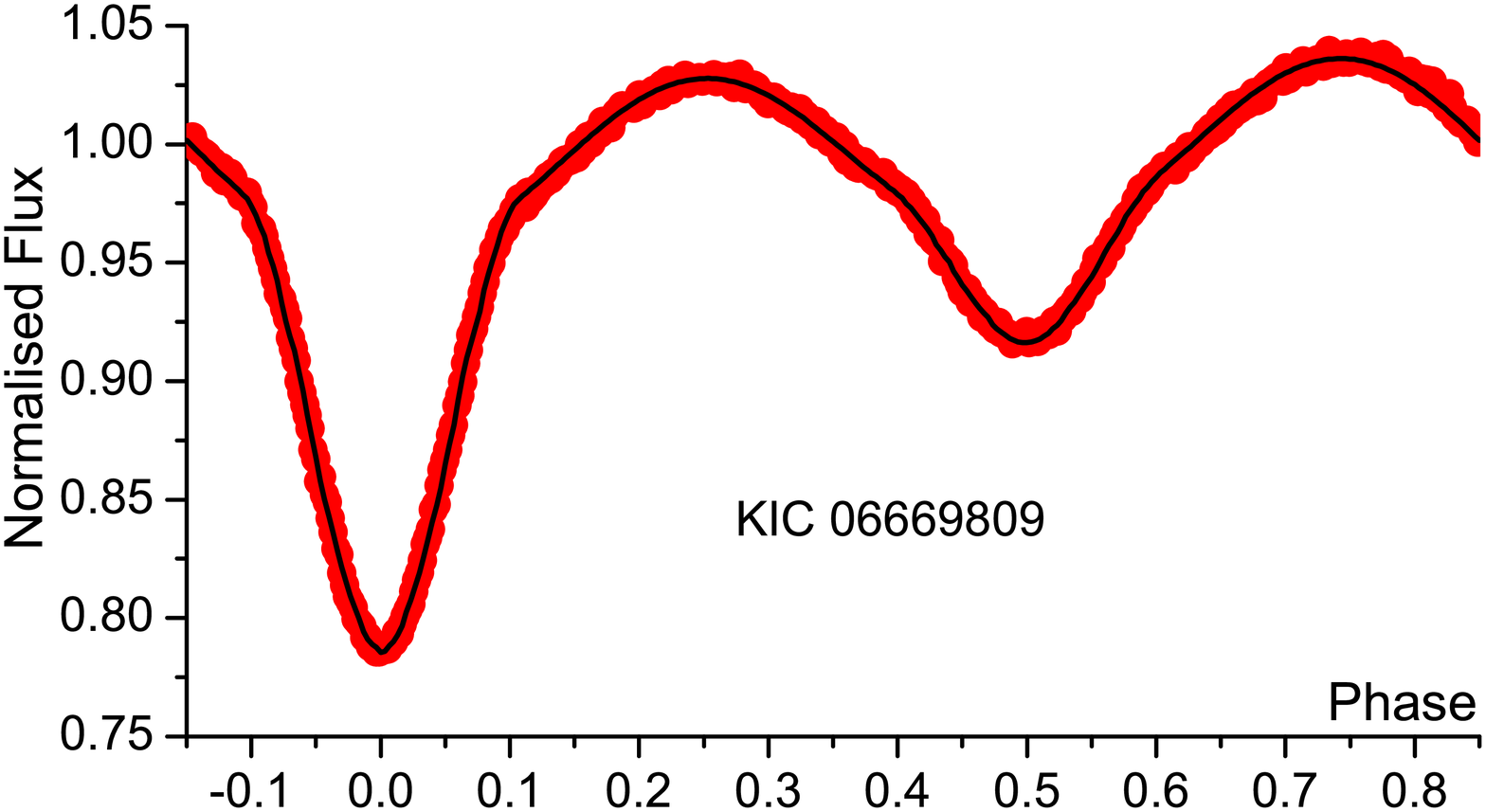}\\
\includegraphics[width=7.5cm]{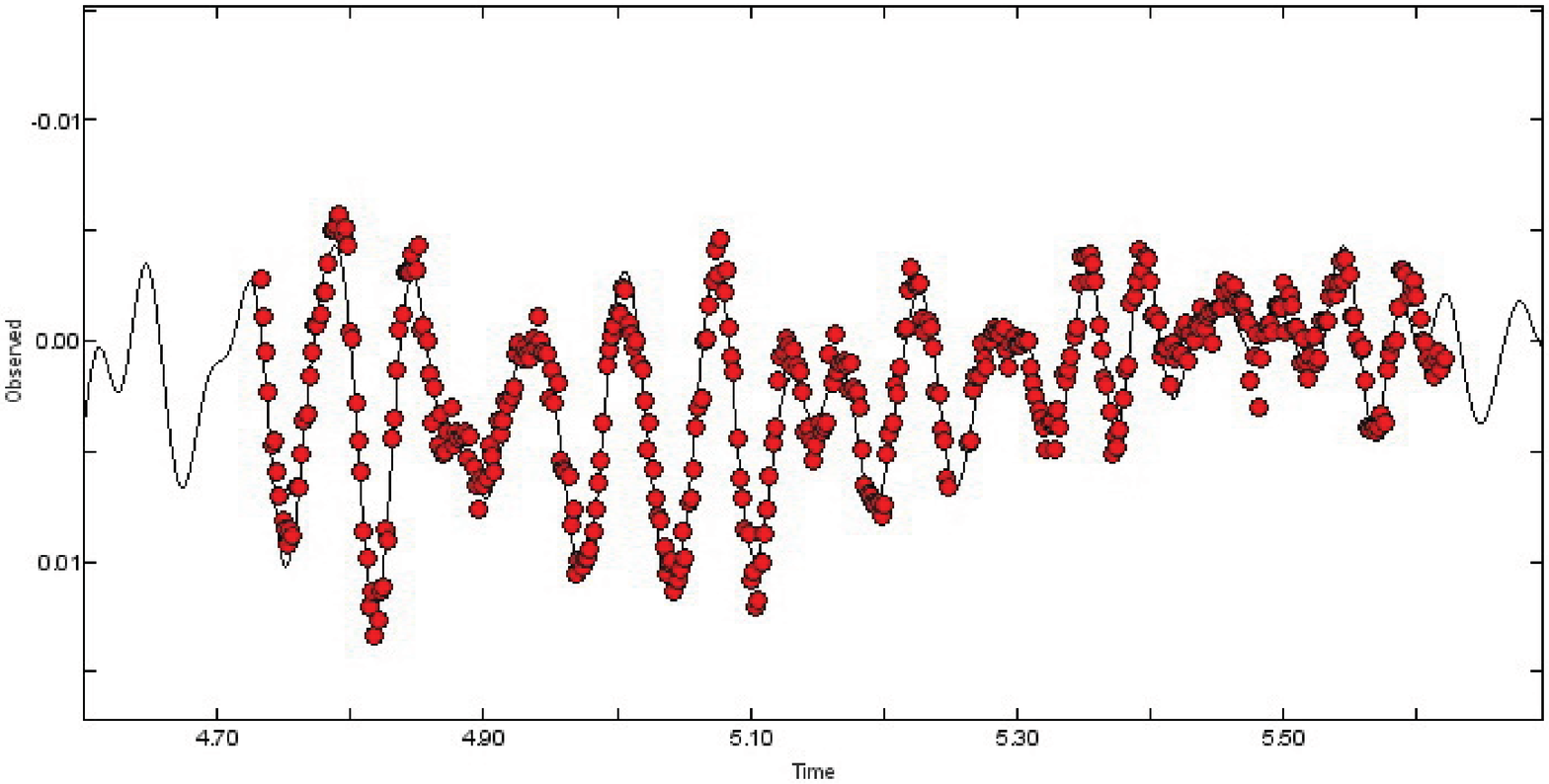}&\includegraphics[width=7.8cm]{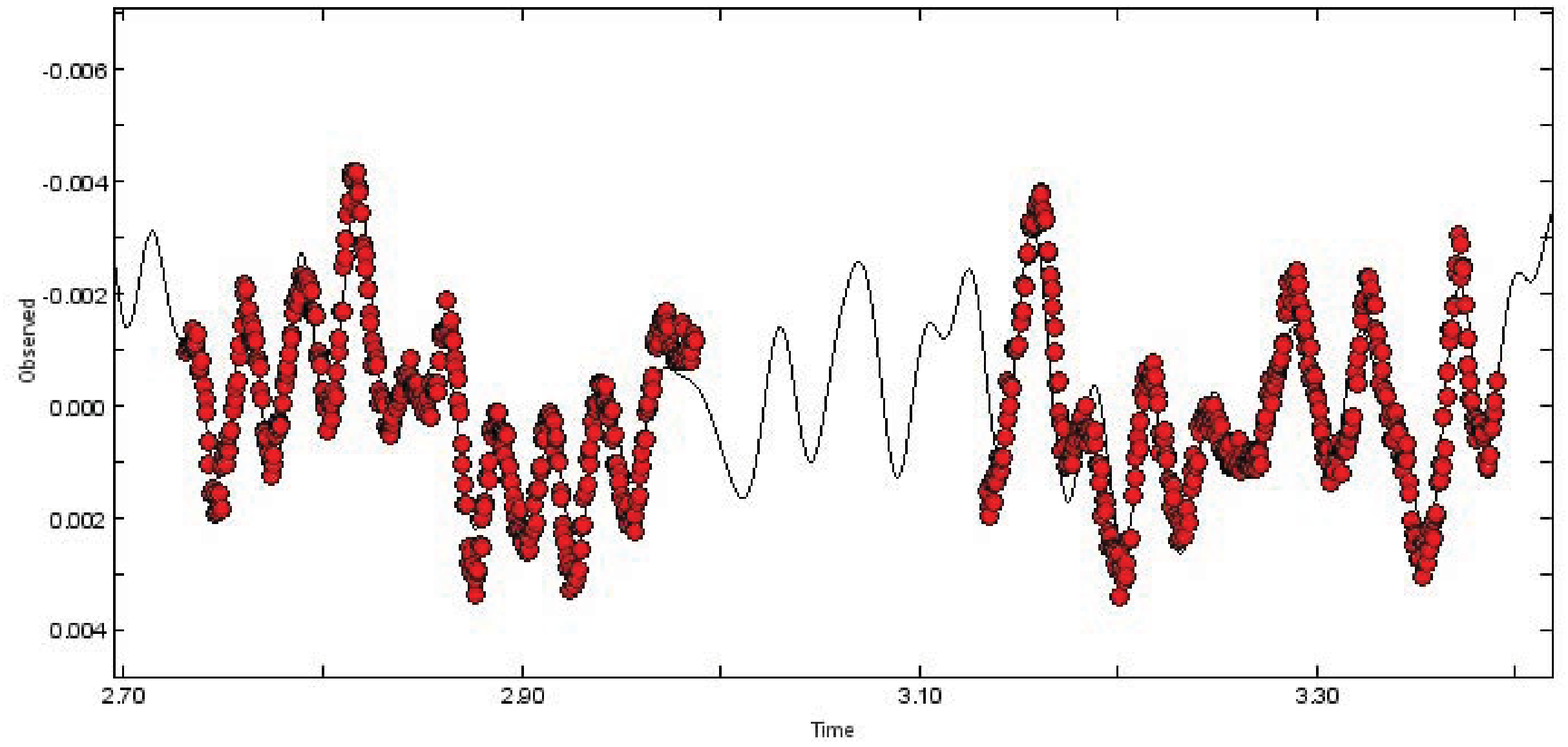}\\
\end{tabular}
\caption{Light curve modelling (up) and Fourier fitting (bottom) on the residuals for KIC~06629588 (left) and KIC~06669809 (right). Red filled circles denote observed points, while theoretical curves are represented by solid lines.}
\label{Fig2}
\end{figure}

Using this database we have found more than 40 eclipsing binaries candidates for including a $\delta$~Sct component. For the present study, we selected two of them for analysis, namely KIC~06629588 and KIC~06669809, and we present preliminary results. For KIC~06629588 we used $\sim$106K data points in a time span of $\sim97$~days. Using standard methods of analysis (see e.g. \cite{LI12}) the system is found to be detached with a mass ratio of $\sim0.83$. We found 10 independent pulsation frequencies for its primary (hotter and more massive) component, with the dominant one to be $f_1=$13.396~c/d. Moreover, 203 more frequencies-combinations of the first 10 were also detected. For KIC~06669809 there are $\sim$32K data points available covering a time span of $\sim30$~days. The binary is identified as a classical Algol system with a mass ratio of $\sim0.42$ and its primary component was found to pulsate in 7 independent oscillation modes ($f_1=$32.564~c/d) and in another 44 combination-frequencies. The light curve modelling as well as the Fourier fit on individual data points after removing the binary solution for both systems are plotted in fig.~\ref{Fig2}. Detailed results for both systems will be presented in a future study.

\noindent
{\bf Acknowledgements:} This work was performed in the framework of PROTEAS project within GSRT's KRIPIS action for A.L., funded by Hellas and the European Regional Development Fund of the European Union under the O.P. Competitiveness and Entrepreneurship, NSRF 2007-2013.


\begin{thebibliography}{}


\bibitem{LI12}
Liakos, A., Niarchos, P., Soydugan, E., Zasche, P.: 2012, MNRAS, 422, 1250.
\bibitem{LI13}
Liakos, A., Niarchos, P.: 2013, Ap\&SS, 341, 123.
\bibitem{LI15}
Liakos, A., Niarchos, P.: 2015, ASPCS, 496, 115.
\bibitem{MK02}
Mkrtichian, D.~E., Kusakin, A.~V., Gamarova, A.~Yu, et al.: 2002, ASPCS, 259, 96.
\bibitem{MK04}
Mkrtichian, D.~E., Kusakin, A.~V., Rodr\'{\i}guez, E., et al.: 2004, A\&A, 419, 1015.
\bibitem{PR11}
Pr{\v s}a, A., Batalha, N., Slawson, R.~W., et al.: 2011, AJ, 141, 83.
\bibitem{SO06a}
Soydugan, E., \.{I}bano\v{g}lu, C., Soydugan, F., et al.: 2006a, MNRAS, 366, 1289.
\bibitem{SO06b}
Soydugan, E., Soydugan, F., Demircan, O., et al.: 2006b, MNRAS, 370, 2013.
\bibitem{ZH13}
Zhang, X.~B., Luo, C.~Q., Fu, J.~N.: 2013, ApJ, 777, 77.








\bibitem{ZH10}
Zhou A.-Y.: 2010, preprint (arXiv:1002.2729v2).



\end{thebibliography}
\end{document}